\documentclass[12pt]{article}
\newcommand{\bra}{\langle}
\newcommand{\ket}{\rangle}
\newcommand{\R}{\mbox{\boldmath $ R $}}
\newcommand{\C}{\mbox{\boldmath $ C $}}
\newcommand{\Z}{\mbox{\boldmath $ Z $}}

\makeatletter
\@addtoreset{equation}{section}
\makeatother

\begin{document} 
\baselineskip 7mm 
\begin{flushright}
{\small
September 9, 2003
\\
revised on October 7, 2003
\\
hep-th/0309091
}
\end{flushright}
\vspace*{8mm}
\begin{center}
{\LARGE \bf 
\baselineskip 9mm 
Quantization on a torus without position operators\footnote{%
To be published in Modern Physics Letters A.}

}
\vspace{8mm}
Shogo Tanimura\footnote{e-mail: tanimura@mech.eng.osaka-cu.ac.jp}
\vspace*{5mm} \\
{\small \it
Graduate School of Engineering, Osaka City University 
\\
Sugimoto, Sumiyoshi-ku, Osaka 558-8585, Japan
}
\vspace{15mm}
\\
{\small Abstract}
\vspace{2mm}
\\
\begin{minipage}[t]{130mm}
\small
\baselineskip 5.6mm 
We formulate quantum mechanics in the two-dimensional torus
without using position operators.
We define an algebra with only momentum operators and shift operators
and construct an irreducible representation of the algebra.
We show that it realizes quantum mechanics of a charged particle
in a uniform magnetic field.
We prove that any irreducible representation of the algebra
is unitarily equivalent to each other.
This work provides a firm foundation for the noncommutative torus theory.
\end{minipage}
\end{center}
\vspace{10mm}
\noindent
{\small
PACS: 
%
02.40.Gh; 
03.65.Ca; 
03.65.Fd; 
04.60.Ds 
%
}

\newpage
\baselineskip 6mm 
\section{Introduction}
Traditionally quantum mechanics in the Euclidean space $ \R^n $
is defined as an irreducible representation 
of the canonical commutation relation (CCR).
Namely, self-adjoint operators 
$ \hat{q}_j, \hat{p}_j (j=1,2,\cdots,n) $
are required to satisfy the CCR
\begin{equation}
	[ \hat{q}_j, \hat{p}_k ] = i \delta_{jk} \hat{1},
	\qquad
	[ \hat{q}_j, \hat{q}_k ] = 
	[ \, \hat{p}_j, \hat{p}_k ] = 
	0
	\label{CCR}
\end{equation}
and the algebra generated by 
$ \{ \hat{q}_j, \hat{p}_j \} $
is represented on the Hilbert space 
$ L^2(\R^n) = \{ \psi : \R^n \to \C \} $,
which consists of square integrable functions,
as
\begin{equation}
	\hat{q}_j \psi(q) = q_j \psi(q),
	\qquad
	\hat{p}_j \psi(q) = -i \frac{\partial}{\partial q_j} \psi(q).
\end{equation}
This is called the Schr{\"o}dinger representation.
The operator $ \hat{q}_j $ is called the position operator
while $ \hat{p}_j $ is called the momentum operator.
It has been proved\cite{Neumann} that
any other irreducible representation of the CCR is unitarily equivalent to
the Schr{\"o}dinger representation
although there still remains subtlety in the argument\cite{CCR1,CCR2}.

When we formulate quantum mechanics in the circle $ S^1 $,
we need to introduce an algebra that is different from the CCR.
A wave function in $ S^1 $ is demanded to be a periodic function such that
$ \psi(q + 2\pi) = \psi(q) $.
However, when it is multiplied by $ q $,
$ q \psi(q) $ is not a periodic function.
Hence the position operator $ \hat{q} $ is not well defined in $ S^1 $.
An alternative formulation of quantum mechanics in $ S^1 $ 
has been given\cite{Mackey}-\cite{Tanimura1}.

When we formulate quantum mechanics in $ \R^2 $
with a uniform background magnetic field $ B $, we have no trouble.
We can write down the Hamiltonian
\begin{equation}
	\hat{H} =
	\frac{1}{2} ( \hat{p}_1 )^2 +
	\frac{1}{2} ( \hat{p}_2 - B \hat{q}_1)^2,
	\label{magnetic Hamiltonian}
\end{equation}
which describe dynamics of a charged particle in the magnetic field.

When we formulate quantum mechanics in the torus $ T^2 $
with the background magnetic field, we have trouble.
The torus has a coordinate system $ (q_1,q_2) \in \R^2 $.
For arbitrary integers $ (n_1, n_2) $,
the point $ (q_1 + 2 \pi n_1, q_2 + 2 \pi n_2 ) $
is to be identified with $ (q_1,q_2) $.
Thus, the operator $ \hat{q}_1 $ 
that appears in (\ref{magnetic Hamiltonian})
is not well defined and, therefore, 
the Hamiltonian (\ref{magnetic Hamiltonian}) is not well defined, either.
So we need to find another formulation 
of quantum mechanics in $ T^2 $ with the magnetic field.

In this Letter we solve this problem.
We define an algebra $ {\cal A} $ without using position operators,
instead using shift operators that characterize symmetry of the torus.
We then show that an irreducible representation of the algebra $ {\cal A} $
realizes quantum mechanics in 
the torus
with the uniform magnetic field.
Moreover, we introduce other operators 
which can be interpreted as a kind of position operators,
and define another algebra $ {\cal A}' $ using them.
We show that $ {\cal A}' $ is actually equivalent to $ {\cal A} $.
So, quantization without position operators is possible
and it reproduces position operators, too.

\section{Main results}
Here we describe definitions and main theorems concisely.
In the rest of Letter we will prove them.
\vspace{12pt}

\noindent
{\it Definition 1:
quantization without position operators.} 
Let $ m $ be an integer.
Assume that there are
self-adjoint operators $ \hat{P}_j $
and
unitary operators $ \hat{V}_j $ $ (j=1,2) $
which satisfy the following relations
\begin{eqnarray}
&&	[ \hat{P}_1, \hat{P}_2 ] = \frac{i m}{2 \pi} \hat{1},
	\label{A relation 1}
	\\
&&	(\hat{V}_1)^m = (\hat{V}_2)^m = \hat{1}, 
	\label{A relation 2}
	\\
&&	\hat{V}_1 \hat{V}_2 = e^{2 \pi i /m} \hat{V}_2 \hat{V}_1,
	\label{A relation 3}
	\\
&&	[ \hat{P}_j, \hat{V}_k ] = 0.
	\label{A relation 4}
\end{eqnarray}
The operators $ \hat{P}_j $ are called
momentum operators while
the operators $ \hat{V}_j $ are called
shift operators.
The algebra $ {\cal A} $ generated by 
$ \{ \hat{P}_1, \hat{P}_2, \hat{V}_1, \hat{V}_2 \} $
is called a magnetic torus algebra without position operators.
\vspace{12pt}

\noindent
{\it Remark: }
The shift operators commute with the momenta, and
therefore they characterize symmetry of the system.
\vspace{12pt}

\noindent
{\it Definition 2:
quantization with position operators.} 
Assume that there are
self-adjoint operators $ \hat{P}_j $
and
unitary operators $ \hat{U}_j $ $ (j=1,2) $
which satisfy the following relations
\begin{eqnarray}
&&	[ \hat{P}_1, \hat{P}_2 ] = \frac{i m}{2 \pi} \hat{1},
	\label{A' relation 1}
	\\
&&	\hat{U}_1 \hat{U}_2 = \hat{U}_2 \hat{U}_1, 
	\label{A' relation 2}
	\\
&&	[ \hat{P}_j, \hat{U}_k ] = \delta_{jk} \hat{U}_k
	\label{A' relation 3}
\end{eqnarray}
The operators $ \hat{U}_j $ are called position operators.
The algebra $ {\cal A}' $ generated by 
$ \{ \hat{P}_1, \hat{P}_2, \hat{U}_1, $ $ \hat{U}_2 \} $
is called a magnetic torus algebra with position operators.
\vspace{12pt}

\noindent
{\it Theorem 1: equivalence of the two quantizations.}
The algebra $ {\cal A}' $ is isomorphic to $ {\cal A} $.
Namely, there is a bijection
$ f : {\cal A}' \to {\cal A} $
such that
$ f( \lambda \hat{A} ) = \lambda f( \hat{A} ) $,
$ f( \hat{A} + \hat{B} ) = f( \hat{A} ) + f ( \hat{B} ) $, 
$ f( \hat{A} \hat{B} ) = f( \hat{A} ) f ( \hat{B} ) $ 
for any $ \lambda \in \C $ and any
$ \hat{A}, \hat{B} \in {\cal A}' $.
\vspace{12pt}

\noindent
{\it Remark: }
The above statement tells that 
any element $ \hat{A} \in {\cal A}' $ can be identified with 
$ f(\hat{A}) \in  {\cal A} $
but it does not say that
$ \hat{U}_j $ is identified with $ \hat{V}_j $.
Even if we write $ f( \hat{U}_j ) $ as $ \hat{U}_j $,
it does not cause confusion.
Then they satisfy the relations
\begin{equation}
	\hat{U}_j \hat{V}_k = e^{2 \pi i \delta_{jk}/m} 
	\hat{V}_k \hat{U}_j.
	\label{UV}
\end{equation}

\noindent
{\it Definition 3: momentum representation of the algebra.}
The space of measurable functions
\begin{equation}
	L^2 (\R) \otimes \C^{|m|} :=
	\left\{
		\phi : \R \times \Z_m \to \C
		\, \Bigg| \,
		\int_{-\infty}^\infty dk
		\sum_{r=1}^{|m|}
		| \phi(k,r) |^2
		< \infty
	\right\}
\end{equation}
becomes a Hilbert space equipped with inner product
\begin{equation}
	\bra \phi_1, \phi_2 \ket
	:=
	\int_{-\infty}^\infty dk
	\sum_{r=1}^{|m|}
	\overline{\phi_1(k,r)} \phi_2(k,r),
	\qquad
	\phi_1, \phi_2 \in L^2 (\R) \otimes \C^{|m|}.
\end{equation}
Here 
$ r $ is an element of $ \Z_m = \Z / m \Z $ and
it is to be understood that $ \phi(k,r+m) = \phi(k,r) $.
Let the algebra $ {\cal A} $ act
on the Hilbert space $ {\cal H} = L^2 (\R) \otimes \C^{|m|} $ as
\begin{eqnarray}
&&	\hat{P}_1 \phi(k,r) = k \, \phi(k,r),
	\label{momentum rep1}
	\\
&&	\hat{P}_2 \phi(k,r) 
	= - \frac{i m}{2 \pi} \frac{\partial}{\partial k} \phi(k,r),
	\label{momentum rep2}
	\\
&&	\hat{V}_1 \phi(k,r) = \phi(k,r-1),
	\label{momentum rep3}
	\\
&&	\hat{V}_2 \phi(k,r) = e^{-2 \pi i r/m} \phi(k,r),
	\label{momentum rep4}
	\\
&&	\hat{U}_1 \phi(k,r) = e^{ 2 \pi i r/m} \phi(k-1,r),
	\label{momentum rep5}
	\\
&&	\hat{U}_2 \phi(k,r) = e^{ 2 \pi i k/m} \phi(k,r-1).
	\label{momentum rep6}
\end{eqnarray}
This is called a momentum representation.
\vspace{12pt}

\noindent
{\it Theorem 2: uniqueness of quantization.}
The momentum representation 
is an irreducible representation of the magnetic torus algebra $ {\cal A} $.
Moreover, any irreducible representation of $ {\cal A} $
is unitarily equivalent to the momentum representation.
\vspace{12pt}

\noindent
{\it Definition 4: position representation of the algebra.}
Let $ L^2_m (T^2) $ denote
the space of measurable functions
$ \psi : \R^2 \to \C $ such that
\begin{eqnarray}
&&	\psi(q_1 + 2 \pi, q_2) = e^{i m q_2} \psi( q_1, q_2 ),
	\label{periodic q1}
	\\
&&	\psi(q_1, q_2 + 2 \pi) = \psi( q_1, q_2 ),
	\label{periodic q2}
	\\
&&	\int_0^{2 \pi} \! dq_1 
	\int_0^{2 \pi} \! dq_2 
	| \psi(q_1,q_2) |^2 < \infty.
\end{eqnarray}
The space $ L^2_m (T^2) $ becomes a Hilbert space
equipped with inner product
\begin{equation}
	\bra \psi_1, \psi_2 \ket
	:=
	\int_0^{2 \pi} \! dq_1 
	\int_0^{2 \pi} \! dq_2 \,
	\overline{\psi_1 (q_1,q_2)} \psi_2 (q_1,q_2),
	\qquad
	\psi_1, \psi_2 \in L^2_m (T^2).
\end{equation}
Let the algebra $ {\cal A} $ act
on the Hilbert space $ \tilde{\cal H} = L^2_m (T^2) $ as 
\begin{eqnarray}
&&	\hat{P}_1 \psi (q_1, q_2)
	=
	-i \frac{\partial}{\partial q_1} \psi (q_1, q_2),
	\label{rep P1}
	\\
&&	\hat{P}_2 \psi (q_1, q_2)
	=
	\left(
		-i \frac{\partial}{\partial q_2} 
		- \frac{m}{2 \pi} q_1
	\right)
	\psi (q_1, q_2),
	\label{rep P2}
	\\
&&	\hat{V}_1 \psi (q_1, q_2)
	=
	e^{i q_2} \psi \left( q_1 - \frac{2 \pi}{m}, q_2 \right),
	\label{rep V1}
	\\
&&	\hat{V}_2 \psi (q_1, q_2)
	=
	\psi \left( q_1, q_2 - \frac{2 \pi}{m} \right),
	\label{rep V2}
	\\
&&	\hat{U}_1 \psi (q_1, q_2) = e^{i q_1} \psi ( q_1, q_2 ),
	\label{rep U1}
	\\
&&	\hat{U}_2 \psi (q_1, q_2) = e^{i q_2} \psi ( q_1, q_2 ).
	\label{rep U2}
\end{eqnarray}
This is called a position representation.
\vspace{12pt}

\noindent
{\it Remark:}
The wave function $ \psi $ 
that satisfies the boundary condition 
(\ref{periodic q1}) and (\ref{periodic q2})
is essentially defined in the torus.
Owing to (\ref{rep U1}) and (\ref{rep U2}), 
it is reasonable to call $ \hat{U}_j $ position operators in the torus.
In (\ref{rep P1}) and (\ref{rep P2})
the momentum operators are represented by the covariant derivative
associated with a uniform magnetic field
\begin{equation}
	B = -i[ \hat{P}_1, \hat{P}_2 ] = \frac{m}{2 \pi}.
\end{equation}
The magnetic flux
$ \int \!\! \int B \, dq_1 \, dq_2 = 2 \pi m $
is quantized as a monopole in a sphere. 
Thus, it is natural to call the algebra $ {\cal A} $
the magnetic torus algebra.
Owing to (\ref{rep V1}) and (\ref{rep V2}), 
it is also reasonable to call $ \hat{V}_j $ shift operators.
They translate the wave function by a finite distance
and they commute with the momentum operators.
Therefore, the shift operators characterize symmetry 
of quantum mechanics in the torus 
as noted in the previous works\cite{Tanimura2,Tanimura3}.
\vspace{12pt}

\noindent
{\it Theorem 3:}
The position representation is an irreducible representation
of the magnetic torus algebra.
\vspace{12pt}

\noindent
{\it Remark:}
As a corollary of theorem 2,
the position representation 
is also unitarily equivalent to the momentum representation.
In this sense, 
quantum mechanics in the torus with the magnetic field
is completely characterized by the algebra without position operators.

\section{Proof of theorem 1}
Assume that the operators
$ \{ \hat{P}_1, \hat{P}_2, \hat{V}_1, \hat{V}_2 \} $
satisfy the defining relations
(\ref{A relation 1})-(\ref{A relation 4})
of the algebra $ {\cal A} $.
If we put
\begin{equation}
	\hat{U}_1 = 
	e^{- 2 \pi i \hat{P}_2 /m } \cdot \hat{V}_2^\dagger,
	\qquad
	\hat{U}_2 =
	e^{2 \pi i \hat{P}_1 /m } \cdot \hat{V}_1,
	\label{U by V}
\end{equation}
it is not difficult to verify that
$ \{ \hat{P}_1, \hat{P}_2, \hat{U}_1, \hat{U}_2 \} $ 
satisfy the defining relations 
(\ref{A' relation 1})-(\ref{A' relation 3})
of the algebra $ {\cal A}' $.
Thus 
we can say that $ {\cal A}' \subset {\cal A} $.

Oppositely, 
by accepting the relations 
of the algebra $ {\cal A}' $ and 
by putting
\begin{equation}
	\hat{V}_1 =
	e^{-2 \pi i \hat{P}_1 /m } \cdot \hat{U}_2,
	\qquad
	\hat{V}_2 = 
	\hat{U}_1^\dagger \cdot e^{- 2 \pi i \hat{P}_2 /m },
	\label{V by U}
\end{equation}
we can show that they satisfy the defining relations
of the algebra $ {\cal A} $.
Thus we can say that $ {\cal A} \subset {\cal A}' $.
Therefore, we conclude that $ {\cal A} = {\cal A}' $.
The identity map
$ \mbox{id} : {\cal A}' \to {\cal A} $
is actually taken to be the isomorphism
$ f : {\cal A}' \to {\cal A} $.
The relations (\ref{UV}) are also easily derived from 
(\ref{A relation 1})-(\ref{A relation 4}) and (\ref{U by V}).

\section{Proof of theorem 2}
It is easy to verify that the equations
(\ref{momentum rep1})-(\ref{momentum rep6})
define a representation of $ {\cal A} $.
Its irreducibility and uniqueness are proved below.

Suppose that we have any irreducible representation space 
of $ {\cal A} $.
The algebra $ {\cal A} $ is generated by
$ \{ \hat{P}_1, \hat{P}_2, \hat{V}_1, \hat{V}_2 \} $.
Two subalgebras generated by
$ \{ \hat{P}_1, \hat{P}_2 \} $ and by
$ \{ \hat{V}_1, \hat{V}_2 \} $ respectively
are mutually commutative.
Thus, it is enough for the proof to consider
$ \{ \hat{P}_1, \hat{P}_2 \} $ and 
$ \{ \hat{V}_1, \hat{V}_2 \} $ separately.
The representation of the whole algebra
is given by a tensor product of representations of each subalgebra.

First, let us concentrate to 
the subalgebra generated by $ \{ \hat{P}_1, \hat{P}_2 \} $.
The defining relation
(\ref{A relation 1})
is isomorphic to the canonical commutation relation (CCR).
It is well known\cite{Neumann} that
any irreducible representation of the CCR is unitarily equivalent
to the Schr{\"o}dinger representation 
(\ref{momentum rep1}) and (\ref{momentum rep2})
over $ L^2 (\R) $.

Second, let us turn to 
the subalgebra $ {\cal B} $
generated by $ \{ \hat{V}_1, \hat{V}_2 \} $.
Take an orthonormal basis 
$ \{ | r \ket $ $ | \, r=1,2,\cdots,|m| \} $
of $ \C^{|m|} $.
If we let $ \{ \hat{V}_1, \hat{V}_2 \} $ act on them as
\begin{equation}
	\hat{V}_1 | r \ket = | r+1 \ket,
	\qquad
	\hat{V}_2 | r \ket = e^{-2 \pi i r/m} | r \ket,
\end{equation}
these actions define a representation of $ {\cal B} $ over $ \C^{|m|} $.
Here it is assumed that $ | r+m \ket = | r \ket $.

We will show that the representation $ ({\cal B}, \C^{|m|}) $ is irreducible.
Suppose that there exists an operator $ \hat{T} $ 
which commutes with any element of $ {\cal B} $.
By taking matrix elements of $ [ \hat{T}, \hat{V}_2 ] = 0 $, we get
\begin{equation}
	e^{-2 \pi i r'/m} \bra r | \hat{T} | r' \ket -
	e^{-2 \pi i r/m} \bra r | \hat{T} | r' \ket 
	=
	( e^{-2 \pi i r'/m} - e^{-2 \pi i r/m} )
	\bra r | \hat{T} | r' \ket 
	= 0.
\end{equation}
Hence, $ \bra r | \hat{T} | r' \ket = 0 $ 
when $ r \ne r' ( \mbox{mod} \, m ) $.
On the other hand, by taking matrix elements of
$ [ \hat{T}, \hat{V}_1 ] = 0 $, we get
\begin{equation}
	\bra r | \hat{T} | r'+1 \ket -
	\bra r-1 | \hat{T} | r' \ket 
	= 0,
\end{equation}
which is equivalent to
\begin{equation}
	\bra r+1 | \hat{T} | r'+1 \ket =
	\bra r | \hat{T} | r' \ket.
\end{equation}
Therefore, the matrix
$ \bra r | \hat{T} | r' \ket $ is diagonal 
and their diagonal elements are equal.
In other words, the operator $ \hat{T} $ is a scalar.
Then the Schur lemma implies that 
the representation $ ({\cal B}, \C^{|m|}) $ is irreducible.

Next we will show that
any other irreducible representation of $ {\cal B} $ is
unitarily equivalent to $ ({\cal B}, \C^{|m|}) $.
Suppose that a Hilbert space $ {\cal E} $ 
provides an irreducible representation of $ {\cal B} $.
Then we define operators
\begin{equation}
	\hat{Q}_r := 
	\frac{1}{|m|} \sum_{q=1}^{|m|} ( e^{2 \pi i r/m} \hat{V}_2 )^q
\end{equation}
that act on $ {\cal E} $.
The set of operators $ \{ \hat{Q}_1, \cdots, \hat{Q}_{|m|} \} $
provides resolution of the identity.
Namely, they satisfy
\begin{equation}
	\hat{Q}_r \hat{Q}_{r'} = \delta_{rr'} \hat{Q}_r, 
	\qquad
	(\hat{Q}_r)^\dagger = \hat{Q}_r,
	\qquad
	\sum_{r=1}^{|m|} \hat{Q}_r = \hat{1}.
\end{equation}
These properties are checked by direct calculations: 
the first one is verified as
\begin{eqnarray}
	\hat{Q}_r \hat{Q}_{r'}
& = &
	\frac{1}{|m|^2}
	\sum_{q,q' = 1}^{|m|}
	( e^{2 \pi i r/m} \hat{V}_2 )^q
	( e^{2 \pi i r'/m} \hat{V}_2 )^{q'}
	\nonumber \\
& = &
	\frac{1}{|m|^2}
	\sum_{q = 1}^{|m|} e^{2 \pi i (r-r') q /m} 
	\sum_{l = 1}^{|m|} e^{2 \pi i r'l/m} ( \hat{V}_2 )^l
	\nonumber \\
& = &
	\frac{1}{|m|^2}
	\, |m| \, \delta_{rr'}
	\sum_{l = 1}^{|m|} ( e^{2 \pi i r'/m} \hat{V}_2 )^l
	\nonumber \\ 
& = &
	\delta_{rr'} \hat{Q}_{r'}.
\end{eqnarray}
Here we put $ q' = l - q $.
The second one is
\begin{equation}
	(\hat{Q}_r)^\dagger 
	=
	\frac{1}{|m|} \sum_{q=1}^{|m|} ( e^{2 \pi i r/m} \hat{V}_2 )^{-q}
	= \hat{Q}_r.
\end{equation}
The third one is
\begin{equation}
	\sum_{r=1}^{|m|} \hat{Q}_r 
	= 
	\frac{1}{|m|}
	\sum_{q=1}^{|m|} ( \hat{V}_2 )^q
	\sum_{r=1}^{|m|} e^{2 \pi i q r/m} 
	= 
	\sum_{q=1}^{|m|} ( \hat{V}_2 )^q
	\, \delta_{q0}
	= 
	\hat{1}.
\end{equation}
It is also easy to see 
\begin{equation}
	\hat{V}_2 \hat{Q}_r = e^{- 2 \pi i r/m} \hat{Q}_r.
\end{equation}
Hence the image of the projection operator $ \hat{Q}_r $
is an eigenspace of $ \hat{V}_2 $ 
associated with the eigenvalue $ e^{- 2 \pi i r/m} $.
Moreover, it is also easy to see 
\begin{equation}
	\hat{V}_1 \hat{Q}_r \hat{V}_1^\dagger = \hat{Q}_{r+1},
\end{equation}
which implies that
all traces of $ \hat{Q}_r $ $ (r=1,\cdots,|m|) $
are equal and hence 
the images of the projection operators $ \hat{Q}_r $ 
have the equal dimensions $ d = \hat{Q}_r $ $ (r=1,\cdots,|m|) $.
Therefore,
the representation $ ( {\cal B}, {\cal E} ) $ can be decomposed 
into $ d $ copies of $ ( {\cal B}, \C^{|m|} ) $.
If $ ( {\cal B}, {\cal E} ) $ is irreducible,
$ d = 1 $ and it is equivalent to $ ( {\cal B}, \C^{|m|} ) $.

Combining above discussions,
we conclude that 
the Hilbert space $ L^2(\R) \otimes \C^{|m|} $ provides
an irreducible representation of the algebra $ {\cal A} $
and that
any other irreducible representation of $ {\cal A} $ is equivalent to it.

\section{Proof of theorem 3}
Our plan of the proof is as follows:
First, 
we will construct another irreducible representation $ l^2(\C) \otimes \C^{|m|} $
of the algebra $ {\cal A} $.
Second, 
we will show that the momentum representation $ L^2(\R) \otimes \C^{|m|} $
is unitarily equivalent to $ l^2(\C) \otimes \C^{|m|} $.
Third,
we will show that the position representation $ L^2_m (T^2) $
is unitarily equivalent to $ l^2(\C) \otimes \C^{|m|} $.
Combining them, we will complete the proof.

The first step: 
If we put
\begin{equation}
	\hat{a} 
	:= \sqrt{\frac{\pi}{|m|}} 
	\left( \hat{P}_1 +i \frac{m}{|m|} \hat{P}_2 \right),
	\qquad
	\hat{a}^\dagger 
	:= \sqrt{\frac{\pi}{|m|}} 
	\left( \hat{P}_1 -i \frac{m}{|m|} \hat{P}_2 \right),
\end{equation}
they satisfy $ [ \hat{a}, \hat{a}^\dagger ] = \hat{1} $.
Let 
$ \{ | n,r \ket $ $ | \, n=0,1,2,\cdots; \, r=1,2,\cdots,|m| \} $
denote a complete orthonormal set of the Hilbert space
$ l^2 (\C) \otimes \C^{|m|} $.
The generators of the algebra $ {\cal A} $ act as
\begin{eqnarray}
&&	\hat{a} | n,r \ket = \sqrt{n} \, | n-1,r \ket,
	\qquad
	\hat{a}^\dagger | n,r \ket = \sqrt{n+1} \, | n+1,r \ket,
	\label{rep osc1}
	\\
&&	\hat{V}_1 | n,r \ket = | n,r+1 \ket,
	\qquad
	\hat{V}_2 | n,r \ket = e^{-2 \pi i r/m} | n,r \ket.
	\label{rep osc2}
\end{eqnarray}
It is well known that $ l^2(\C) $ provides an irreducible representation
of the CCR.
Let us call this representation an oscillator representation.

The second step: 
Putting 
$ \phi_{n,r} (k,r') 
= \phi_{n} (k) \, \delta_{r r'}
= \bra k,r' | n,r \ket $
and combining 
(\ref{momentum rep1}), (\ref{momentum rep2})
with
(\ref{rep osc1}),
we get the set of equations
\begin{eqnarray}
&&	\sqrt{\frac{\pi}{|m|}} 
	\left( 
		k +
		\frac{|m|}{2 \pi} \frac{\partial}{\partial k} 
	\right)
	\phi_{n} (k)
	= \sqrt{n} \, \phi_{n-1} (k),
	\\
&&	\sqrt{\frac{\pi}{|m|}} 
	\left( 
		k -
		\frac{|m|}{2 \pi} \frac{\partial}{\partial k} 
	\right)
	\phi_{n} (k)
	= \sqrt{n+1} \, \phi_{n+1} (k).
\end{eqnarray}
The normalization condition
$ \int_{-\infty}^\infty dk | \phi_n(k) |^2 = 1 $
is also imposed on them. The solution is uniquely given as
\begin{equation}
	\phi_n (k) 
	=
	\frac{1}{\sqrt{2^n n!}}
	\left( \frac{2}{|m|} \right)^{\frac{1}{4}}
	e^{-\pi k^2 / |m| } \,
	H_n 
	\left(
		\sqrt{\frac{2 \pi}{|m|}} \, k
	\right),
	\label{nk function}
\end{equation}
where 
$ 
	H_n ( \xi ) 
	= (-1)^n \,
	e^{\xi^2} \frac{d^n}{d \xi^n} e^{-\xi^2}
$ is the $ n $-th Hermite polynomial.
The set of functions
$ \{ \phi_{n,r} | \, n=0,1,2,\cdots; \, r=1,2,\cdots,|m| \} $
constitutes a complete orthonormal set of $ L^2(\R) \otimes \C^{|m|} $.
Then the linear map
\begin{equation}
	{\mit\Gamma}_1 : 
	l^2(\C) \otimes \C^{|m|} \to L^2(\R) \otimes \C^{|m|},
	\qquad
	\sum_{n=0}^\infty \sum_{r=1}^{|m|} c_{n,r} | n,r \ket
	\mapsto
	\sum_{n=0}^\infty \sum_{r=1}^{|m|} c_{n,r} \phi_{n,r} (k,r')
\end{equation}
becomes a unitary transformation that bridges
the oscillator representation and the momentum representation.

The third step: 
Putting 
$ \psi_{n,r} (q_1,q_2) 
= \bra q_1,q_2 | n,r \ket $
and combining 
(\ref{rep P1})-
(\ref{rep V2})
with
(\ref{rep osc1}), (\ref{rep osc2}),
we get the set of equations
\begin{eqnarray}
&&	\sqrt{\frac{\pi}{|m|}} 
	\left( 
		-i \frac{\partial}{\partial q_1} 
		+ \frac{m}{|m|}
		\bigg(
			 \frac{\partial}{\partial q_2} 
			-i \frac{m}{2 \pi} q_1
		\bigg)
	\right)
	\psi_{n,r} (q_1,q_2)
	= \sqrt{n} \, \psi_{n-1,r} (q_1,q_2),
	\\
&&	\sqrt{\frac{\pi}{|m|}} 
	\left( 
		-i \frac{\partial}{\partial q_1} 
		- \frac{m}{|m|}
		\bigg(
			 \frac{\partial}{\partial q_2} 
			-i \frac{m}{2 \pi} q_1
		\bigg)
	\right)
	\psi_{n,r} (q_1,q_2)
	= \sqrt{n+1} \, \psi_{n+1,r} (q_1,q_2),
	\\
&&	e^{i q_2} \psi_{n,r} \left( q_1 - \frac{2 \pi}{m}, q_2 \right)
	=
	\psi_{n,r+1} (q_1,q_2),
	\\
&&	\psi_{n,r} \left( q_1, q_2 - \frac{2 \pi}{m} \right)
	=
	e^{-2 \pi i r/m} \psi_{n,r} (q_1,q_2).
\end{eqnarray}
The boundary condition
(\ref{periodic q1}), (\ref{periodic q2})
and the normalization condition
$ \int_0^{2 \pi} dq_1 dq_2 $ $ | \psi_{n,r} (q_1, $ $ q_2) |^2 $ $ = 1 $
are imposed on them. 
By a tedious calculation\cite{Tanimura3} we get the unique solution 
\begin{eqnarray}
	\psi_{n,r} (q_1,q_2)
& = &	
	\frac{i^n}{\sqrt{2^n n!}} ( 2|m| )^{\frac{1}{4}}
	\sum_{l= -\infty}^\infty 
	e^{i (ml+r) q_2 }
	\nonumber \\ &&
	e^{-|m| \{ q_1 - 2 \pi (ml+r)/m \}^2/ (4 \pi) } \,
	H_n
	\left[
		\sqrt{\frac{|m|}{2 \pi}}
		\{
			q_1 - 2 \pi (ml+r)/m 
		\}
	\right].
	\label{nq function}
\end{eqnarray}
The set of functions
$ \{ \psi_{n,r} | \, n=0,1,2,\cdots; \, r=1,2,\cdots,|m| \} $
constitutes a complete orthonormal set of $ L^2_m (T^2) $.
Then the linear map
\begin{equation}
	{\mit\Gamma}_2 : 
	l^2(\C) \otimes \C^{|m|} \to L^2_m (T^2),
	\qquad
	\sum_{n=0}^\infty \sum_{r=1}^{|m|} c_{n,r} | n,r \ket
	\mapsto
	\sum_{n=0}^\infty \sum_{r=1}^{|m|} c_{n,r} \psi_{n,r} (q_1,q_2)
\end{equation}
becomes a unitary transformation that bridges
the oscillator representation and the position representation.

The final step: 
It is obvious that 
the combined map 
$ {\mit\Gamma} = {\mit\Gamma}_2 \circ {\mit\Gamma}_1^{-1} $
is a unitary transformation
which transforms 
the momentum representation
to 
the position representation.

\section{Concluding remarks}
We leave some remarks in order.
The unitary transformation
$ {\mit\Gamma} : 
L^2(\R) \otimes \C^{|m|} 
\to 
L^2_m (T^2) $
is concretely given\cite{Tanimura3} as
\begin{equation}
	\psi (q_1,q_2)
	=
	\sum_{l= -\infty}^\infty 
	\sum_{r=1}^{|m|}
	\int_{-\infty}^\infty dk \,
	e^{i k ( q_1 - 2 \pi (ml+r)/m ) + i (ml+r) q_2 } \,
	\phi (k,r).
\end{equation}
The transforming function
\begin{equation}
	\chi_{k,r} (q_1,q_2)
	= \bra q_1,q_2 | k,r \ket
	= 
	\sum_{l= -\infty}^\infty 
	e^{i k ( q_1 - 2 \pi (ml+r)/m ) + i (ml+r) q_2 }
	\label{transforming function}
\end{equation}
is a formal solution of the equations
\begin{eqnarray}
&&	
	-i \frac{\partial}{\partial q_1} \chi_{k,r} (q_1, q_2)
	= 
	k \, \chi_{k,r} (q_1, q_2),
	\label{problem:1} \\
&&	
	\left(
		-i \frac{\partial}{\partial q_2} 
		- \frac{m}{2 \pi} q_1
	\right)
	\chi_{k,r} (q_1, q_2)
	= 
	\frac{i m}{2 \pi} \frac{\partial}{\partial k} \,
	\chi_{k,r} (q_1, q_2),
	\label{problem:2} \\
&&	
	e^{i q_2} \, \chi_{k,r} \left( q_1 - \frac{2 \pi}{m}, q_2 \right)
	= 
	\chi_{k,(r+1)} (q_1, q_2),
	\label{problem:3} \\
&&	
	\chi_{k,r} \left( q_1, q_2 - \frac{2 \pi}{m} \right)
	=
	e^{-2 \pi i r/m} \, \chi_{k,r} (q_1, q_2),
	\label{problem:4} \\
&&	\chi_{k,(r+m)} (q_1, q_2)
	=
	\chi_{k,r} (q_1, q_2).
	\label{problem:5}
\end{eqnarray}
However, the infinite sum with respect to $ l $ in 
(\ref{transforming function}) 
does not converge
and hence $ \chi_{k,r} $ is called a `formal' solution.
It is a natural consequence of the fact that
the operator $ \hat{P}_1 $ has a continuous spectrum 
$ - \infty < k < \infty $
and the generally-known fact that
there is no normalizable eigenfunction for a continuous spectrum.
It is to be noted that
the spectrum of the momentum is not discrete 
although the torus is a compact space.

We shall mention another representation of the algebra $ {\cal A} $.
If we replace (\ref{rep P1})-
(\ref{rep U2}) by
\begin{eqnarray}
&&	\rho_\alpha( \hat{P}_1 ) \psi (q_1,q_2)
	=
	\left(
		-i \frac{\partial}{\partial q_1} - \alpha_1
	\right)
	\psi (q_1,q_2)
	\label{rep P1 alpha} \\
&&	\rho_\alpha( \hat{P}_2 ) \psi (q_1,q_2)
	=
	\left(
		-i \frac{\partial}{\partial q_2} 
		- \frac{m}{2 \pi} q_1 - \alpha_2
	\right)
	\psi (q_1,q_2),
	\\
&&	\rho_\alpha( \hat{U}_1 ) \psi (q_1, q_2) 
	= e^{i ( q_1 + 2 \pi \alpha_2/m) } \psi ( q_1, q_2 ),
	\\
&&	\rho_\alpha( \hat{U}_2 ) \psi (q_1, q_2) 
	= e^{i ( q_2 - 2 \pi \alpha_1/m) } \psi ( q_1, q_2 ),
	\label{rep U2 alpha} 
\end{eqnarray}
with leaving $ \rho_\alpha ( \hat{V}_j ) = \hat{V}_j $,
we get another irreducible representation.
Here $ ( \alpha_1, \alpha_2 ) \in \R^2 $ are arbitrary parameters,
which cause the Aharonov-Bohm effect.
However, it is again unitarily equivalent to (\ref{rep P1})-
(\ref{rep U2}) as seen below.
If we introduce 
\begin{equation}
	\hat{S}_\alpha \psi (q_1,q_2)
	= 
	e^{i \alpha_1 q_1 } \,
	\psi \!
	\left(
		q_1 + \frac{2 \pi}{m} \alpha_2, 
		q_2 - \frac{2 \pi}{m} \alpha_1
	\right),
	\label{S}
\end{equation}
then $ \hat{S}_\alpha $ is a unitary operator acting on $ L^2_m (T^2) $
and it satisfies
\begin{equation}
	\hat{S}_\alpha \hat{P}_j \hat{S}_\alpha^{-1}
	= 
	\rho_\alpha( \hat{P}_j ),
	\qquad
	\hat{S}_\alpha \hat{V}_j \hat{S}_\alpha^{-1}
	= 
	\rho_\alpha( \hat{V}_j ),
	\qquad
	\hat{S}_\alpha \hat{U}_j \hat{S}_\alpha^{-1}
	= 
	\rho_\alpha( \hat{U}_j ).
\end{equation}
Hence we can conclude that
the representation (\ref{rep P1 alpha})-(\ref{rep U2 alpha})
is unitarily equivalent to
to the original one defined by (\ref{rep P1})-(\ref{rep U2}).
It is to be noted that
the operator (\ref{S}) makes sense only when $ m $ is nonzero.
If $ m = 0 $ and $ \alpha_j $ is not an integer, 
there is no unitary operator which transforms
$ \hat{P}_j $ to $ \rho_\alpha( \hat{P}_j ) $.

In our formulation,
the Hamiltonian (\ref{magnetic Hamiltonian}), 
which was not well defined in $T^2 $,
is correctly replaced by
\begin{equation}
	\hat{H} =
	\frac{1}{2} ( \hat{P}_1 )^2 +
	\frac{1}{2} ( \hat{P}_2 )^2.
\end{equation}
Its spectrum yields the Landau levels
and the eigenfunctions have been written down as
(\ref{nk function}) and (\ref{nq function}).
We close our discussion by mentioning that
the magnetic torus algebra,
which is defined by the relations (\ref{A relation 1})-(\ref{UV}),
is equivalent to the so-called noncommutative torus algebra\cite{Connes}.
However, 
our formulation has clear definitions and an reasonable interpretation.
Moreover, it reveals the new fact
that quantization in a topologically nontrivial space
is possible without resorting to position operators.
These points are advantages of our formulation.

Let us put a final comment;
in our formulation the gauge field is assumed to be a fixed background.
It is desirable to extend the formulation
to field theory for more realistic application.
For this point, Ho and Hosotani\cite{Hosotani} have examined
in detail
the translation symmetry of the Chern-Simons gauge system in a torus.

\section*{Acknowledgments}
I have learned much about quantum mechanics in manifolds
from collaborations with I. Tsutsui and M. Sakamoto.
I wish to thank them.
M. Takeori called my attention to the uniqueness problem of quantization
and 
my interest in the problem grew from discussions with him.
M. Nakahara invited me 
to write an introductory article on topology and quantum mechanics.
When I was preparing the article 
and re-examining my previous work on quantum mechanics in the torus,
I reached the conclusion that I present here.
I thank Takeori and Nakahara for making opportunity of this work.
I am grateful to people who have been encouraging me at various aspects.
Discussions during 
the workshop YITP-W-03-07 on ``QFT2003'' 
at the Yukawa Institute for Theoretical Physics 
were also useful to complete this work.


\baselineskip 5.8mm 


\begin{thebibliography}{99}
\bibitem{Neumann}
	J. von Neumann, Ann, Math. {\bf 104} (1931) 570.
\bibitem{CCR1}
	H. Reeh,
	{\sl ``A remark concerning canonical commutation relations''},
	J. Math. Phys. {\bf 29} (1988) 1535. 
\bibitem{CCR2}
	A. Arai,
	{\sl ``Momentum operators with gauge potentials, local quantization of
	magnetic flux, and representation of canonical commutation relations''},
	J. Math. Phys. {\bf 33} (1992) 3374.
\bibitem{Mackey}
	G. W. Mackey,
	{\sl Induced representations of groups and quantum mechanics},
	(Benjamin, New York 1968).
\bibitem{Isham}
	C. J. Isham, 
	{\sl ``Topological and global aspects of quantum theory''},
	in {\sl Relativity, Groups and Topology II}, 
	B. S. DeWitt and R. Stora, Eds
	(North-Holland, Amsterdam, 1984) 1059. 
\bibitem{OK}
	Y. Ohnuki, S. Kitakado,
	{\sl ``Fundamental algebra for quantum mechanics on $ S^D $
	and gauge potentials''},
	J. Math. Phys. {\bf 34} (1993) 2827.
\bibitem{Tanimura1}
	S. Tanimura, 
	{\sl ``Gauge field, parity and uncertainty relation
	of quantum mechanics on $ S^1 $''}, 
	Prog. Theor. Phys. {\bf 90} (1993) 271; arXiv hep-th/9306098.
\bibitem{Tanimura2}
	S. Tanimura,
	{\sl ``Magnetic translation groups 
	in $ n $-dimensional torus and their representations''},
	J. Math. Phys. {\bf 43}, 5926 (2002); arXiv hep-th/0205053.
\bibitem{Tanimura3}
	M. Sakamoto,
	S. Tanimura,
	{\sl ``An extension of Fourier analysis for the $ n $-torus 
	in the magnetic field and its application to spectral analysis 
	of the magnetic Laplacian''},
	J. Math. Phys. (in press); arXiv hep-th/0306006.
\bibitem{Connes}
	A. Connes, M. R. Douglas, A. Schwarz,
	{\sl ``Noncommutative Geometry and Matrix Theory: 
	Compactification on Tori''},
	JHEP {\bf 9802}, 003 (1998); arXiv hep-th/9711162.
\bibitem{Hosotani}
	C-L. Ho, Y. Hosotani,
	{\sl ``Operator algebra in Chern-Simons theory on a torus''},
	Phys. Rev. Lett. {\bf 70} 1360 (1993); 
	arXiv hep-th/9210103.
\end{thebibliography}
\end{document}